\documentclass[]{aastex631}

\begin{document}

\title{Exploring the physical properties of the $\gamma$ Dor binary star RX Dra with photometry and asteroseismology}

\correspondingauthor{Wen-Ping Liao}
\email{liaowp@ynao.ac.cn}	
\author{Li Ping}
\affiliation{Yunnan Observatories, Chinese Academy of Sciences (CAS), 650216 Kunming, P.R. China}
\affiliation{University of Chinese Academy of Sciences, No.1 Yanqihu East Rd, Huairou District, Beijing, P.R.China 101408}

\author{Liao Wen-Ping}
\affiliation{Yunnan Observatories, Chinese Academy of Sciences (CAS), 650216 Kunming, P.R. China}
\affiliation{University of Chinese Academy of Sciences, No.1 Yanqihu East Rd, Huairou District, Beijing, P.R.China 101408}

\author{Sun Qi-Bin}
\affiliation{Yunnan Observatories, Chinese Academy of Sciences (CAS), 650216 Kunming, P.R. China}
\affiliation{University of Chinese Academy of Sciences, No.1 Yanqihu East Rd, Huairou District, Beijing, P.R.China 101408}

\author{Li Min-Yu}
\affiliation{Yunnan Observatories, Chinese Academy of Sciences (CAS), 650216 Kunming, P.R. China}
\affiliation{University of Chinese Academy of Sciences, No.1 Yanqihu East Rd, Huairou District, Beijing, P.R.China 101408}


\begin{abstract}

We model the TESS light curve of the binary system RX Dra, and also first calculate a lot of theoretical models to fit the g-mode frequencies previously detected from the TESS series of this system. The mass ratio is determined to be $q$=0.9026$^{+0.0032}_{-0.0032}$. We newly found that there are 16 frequencies (F1-F7, F11-F20)  identified as dipole g-modes, two frequencies (F21, F22) identified as quadrupole g-modes, and another two frequencies (F23, F24) identified as g-sextupole modes, based on these model fits. The primary star is newly determined to be a $\gamma$ Dor pulsator in the main-sequence stage with a rotation period of about 5.7$^{+0.7}_{-0.2}$ days, rotating slower than the orbital motion. The fundamental parameters of two components are firstly estimated as follows: $M_1$=1.53$^{+0.00}_{-0.17}$ M $_{\odot}$, $M_2$= 1.38$^{+0.18}_{-0.00}$ M $_{\odot}$, $T_1$=7240$^{+490}_{-44}$ K, $T_2$=6747$^{+201}_{-221}$ K, $R_1$=1.8288$^{+0.0260}_{-0.0959}$  R $_{\odot}$, $R_2$= 1.3075$^{+0.0450}_{-0.2543}$ R $_{\odot}$, $L_1$=8.2830$^{+1.8015}_{-0.6036}$ L $_{\odot}$   and $L_2$= 3.4145$^{+0.1320}_{-0.1843}$ L $_{\odot}$. Our results show that the secondary star is in the solar-like pulsator region of the H-R diagram, indicating that it could be a pulsating star of this type. Finally, the radius of the convective core of the primary star is estimated to be about 0.1403$^{+0.0206}_{-0.0000}$ R$_{\odot}$.

\end{abstract}

\keywords{Classical Novae (251) --- Ultraviolet astronomy(1736) --- History of astronomy(1868) --- Interdisciplinary astronomy(804)}


\section{Introduction} \label{sec:intro}

There are a number of very spectacular eclipsing binaries that consist of at least one pulsating star. By studying the pulsation frequencies that occur in the inner part of the oscillating binary components, such systems allow us to probe the interior of the stars. The determination of precise fundamental stellar parameters and information about the interior of the star are certainly important and provide us with powerful tools to test the validity of stellar structure models and to compare evolutionary models for single stars and stars in binary systems  \citep{2022MNRAS.510.1413K}.

$\gamma$ Doradus ($\gamma$ Dor) stars are a class of main-sequence variables that were identified about two decades ago \citep{1993MNRAS.263..781K,1994MNRAS.270..439M,1994MNRAS.270..905B,1995IBVS.4195....1K}. This type of variable has sinusoidal light curves with amplitudes ranging from a few mmag to a few percent, and pulsates with multiple photometric periods between 0.3 and 3 days \citep{2005AJ....129.2815H}. Their spectral types range from A 7 to F 5 (6900 $\leq$ T $\leq$ 7400 K) with luminosity classes IV-V or V \citep{1999PASP..111..840K}. The instability strip of $\gamma$ Dor stars is very narrow on the Hertzsprung-Russell (H-R) diagram. \cite{1999MNRAS.309L..19H} had reported that $\gamma$ Dor stars  lie in the lower part of the classical instability band, which partially overlaps the red edge of the $\delta$ Scuti ($\delta$ Sct) instability band. $\gamma$ Dor stars pulsate in gravity (g) modes, which are thought to be driven by the convective blocking mechanism \citep{2000ApJ...542L..57G,2005A&A...434.1055G,2005A&A...435..927D}.In the last few years, asteroseismic analyses of $\gamma$ Dor stars with g-mode pulsations have allowed us to place constraints on a number of phenomena. These include convective core boundary mixing \citep{2019ApJ...881...86W,2019A&A...628A..76M,2021A&A...650A..58M}, near-core stellar rotation \citep{2013MNRAS.429.2500B,2018A&A...618A..47C,2020A&A...635A.106T,2020A&A...644A.138T,2021MNRAS.503.5894S}, extra envelope mixing \citep{2021A&A...650A..58M,2021NatAs...5..715P}, and magnetic filed \citep{2019A&A...627A..64P, 2020A&A...638A.149V,2022MNRAS.512L..16L}. 

Asteroseismology and binarity can be combined for objects showing both eclipses and pulsations. These systems have the potential to place stringent constraints on stellar theory \citep{2019A&A...628A..25J}. A large number of $\gamma$ Dor pulsating stars are known in EBs. \cite{2019RAA....19....1Q} presented the LAMOST data on 168 $\gamma$ Dor stars, and 33 of them are identified as candidates for binary or multiple systems. \cite{2019A&A...630A.106G} reported a sample of 115 $\gamma$ Dor variables in EBs after a systematic search of the Kepler EB catalogue \citep{2016AJ....151...68K}. After performing an asteroseismic evaluation of these targets, \cite{2020MNRAS.497.4363L} reported the detection of g-mode period-spacing patterns for 35 of them. An independent systematic search and asteroseismic evaluation of the Kepler EB catalogue was carried out by \cite{2020A&A...643A.162S}, which resulted in a different sample of 93 $\gamma$ Dor variables in EBs, with period-spacing patterns detected for seven of them. So far, asteroseismic modelling has only been achieved for a few targets: KIC 10080943 \citep{2016A&A...592A.116S, 2019A&A...628A..25J}, KIC 10486425 \citep{2018ApJ...865..115Z}, KIC 7385478 \citep{2019ApJ...882L...5G}, KIC 9850387 \citep{2020ApJ...895..124Z,2021A&A...648A..91S}, and CoRot 100866999 \citep{2019ApJ...887..253C}.

RX Dra (TIC 377190161) has been known for over a century as an eclipsing binary system with an orbital period of 3.78 days. The first photometric analysis was performed by \cite{1913ApJ....38..158S}. However, it has received very little attention since then. \cite{2006MNRAS.370.2013S} put RX Dra on a list of EBs that are candidates for hosting pulsating stars. \cite{2022MNRAS.515.2755S} had performed a frequency analysis for RX Dra and obtained 24 frequencies where 4 frequencies form combinations within 3$\sigma$. However, the pulsations of the RX Dra cannot be attributed to a specific component. In addition, the measured g-mode frequencies are too sparse to detect clear g-mode period spacing patterns, which would have allowed them to perform pulsation mode identification. In this work, we extend the work of \cite{2022MNRAS.515.2755S} and perform a comprehensive asteroseismic modelling and potometric analysis for RX Dra.

The paper is organised as follows. In the section \ref{OC}, we presented the O-C analysis for RX Dra. In the section \ref{W-D}, the photometric model is studied. In the section \ref{MESA}, we carried out the model identification for the pulsating frequencies and the foundational parameters are computed using the MESA code. In the final section, the main discussions and conclusions are given.

\begin{figure}
	\centering 
	\includegraphics[width=0.85\textwidth]{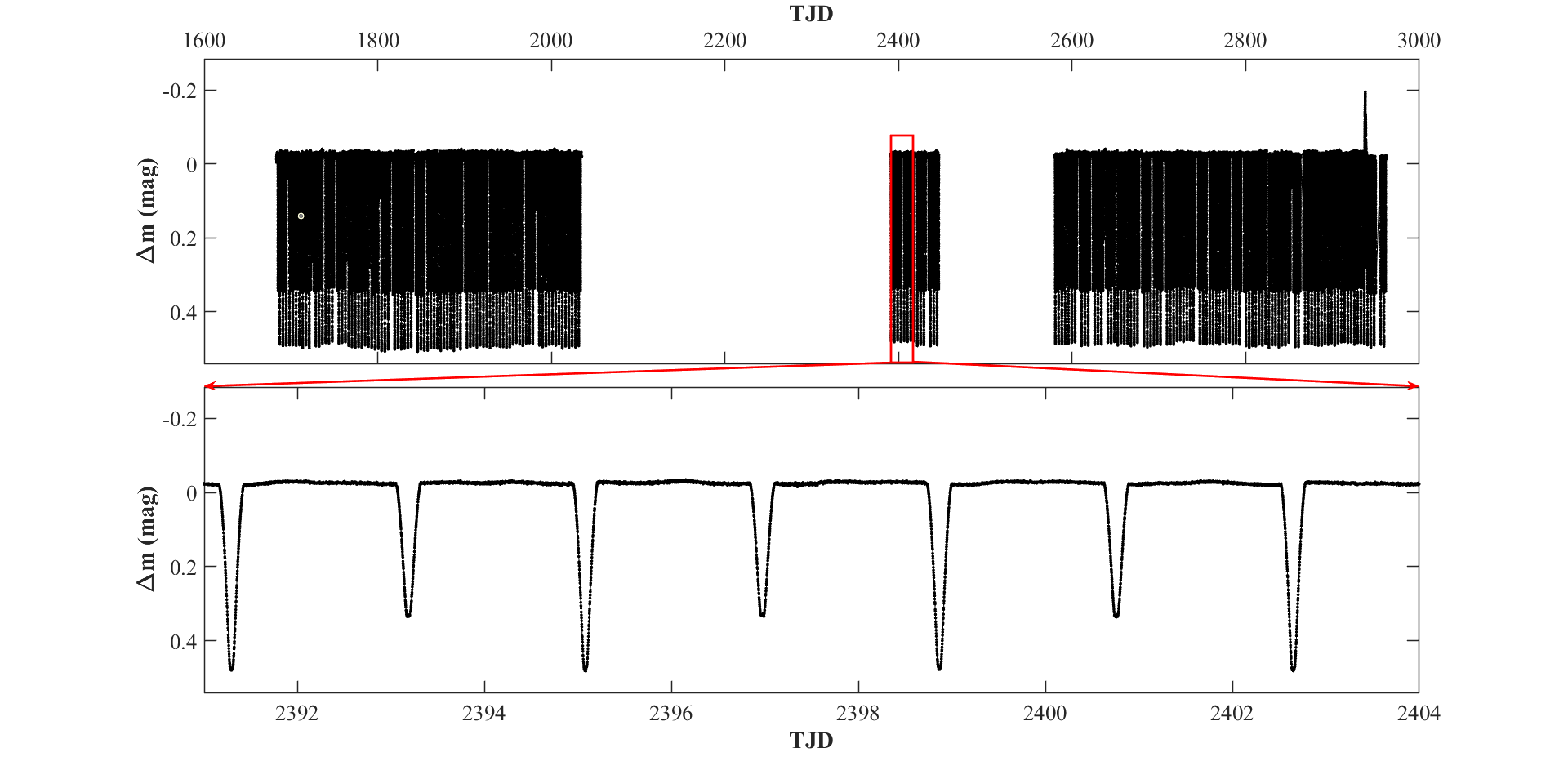}
	\caption{The light curves of RX Dra after processing. Upper panel: all light curves of 32 sectors. Lower panel: a part of the light curve is shown in an enlarged drawing.} 
	\label{fig:figure1}%
\end{figure} 

\section{The analysis of changes in orbital period}
\label{OC}
RX Dra was observed by the TESS mission (Ricker et al. 2015) from TJD 1683.3549 to 3367.4813 (hereafter TJD=BJD-2457000.0) and 32 sectors of light curves were obtained with an exposure time of 120s. These light curves can be downloaded from the MAST\footnote{MAST://archive.stsci.edu/mast.htm} database. Two steps were taken to process the original light curves (Shi et al. 2021): (i) flux to magnitude conversion, (ii) equalisation of the light curves by subtracting the average of each sector. The light curves after processing are shown in Fig.\ref{fig:figure1}, where a part of the light curve is shown in detail with a long period oscillation of small amplitude is clearly visible in the lower panel. 

\begin{figure}
	\centering 
	\includegraphics[width=0.85\textwidth]{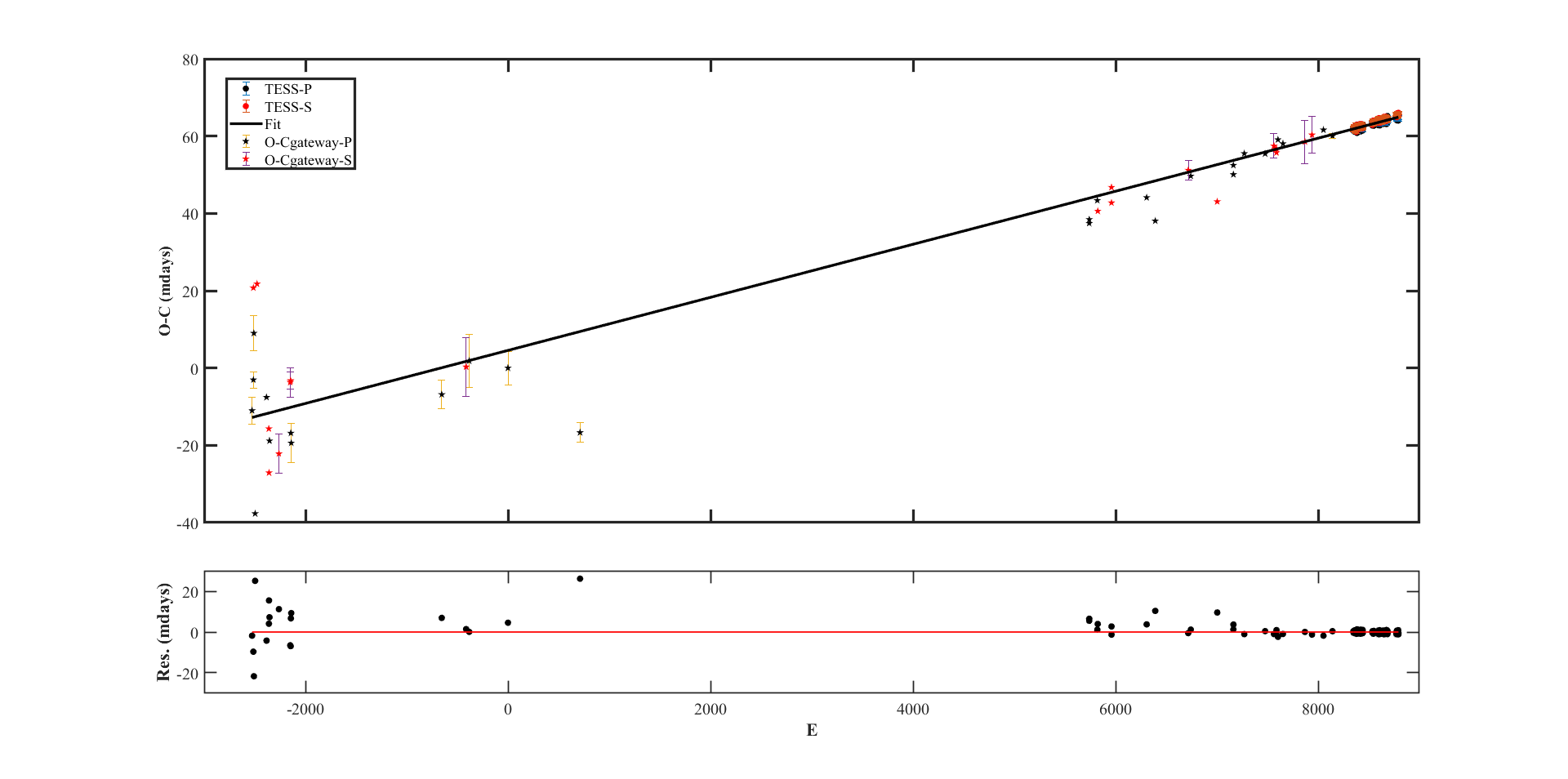}	
	\caption{The O-C plots of RX Dra. The black and red dots indicate these data from TESS, while the black and red pentagons represent data from O-C gateway. The black solid line refer to the revised ephemeris.} 
	\label{fig:figure2}%
\end{figure}

In order to analyse the changes in the orbital period of RX Dra over a long period of time, 211 primary eclipsing times and 210 secondary eclipsing times were obtained from 32 TESS light curve sectors by using parabola fit and 27 primary eclipsing times and 20 secondary eclipsing times were collected from the O-C gateway\footnote{O-C gateway://http://var2.astro.cz/ocgate/} that observed by (Seares et al. 1907; Shapley, 1913; Lavrov et al. 1985). A sample of eclipsing times are listed in Table \ref{tab:tab 1}. Using the primary eclipsing time and the orbital period obtained by \cite{1913ApJ....38..158S}, the linear ephemeris is derived as
\begin{eqnarray}\label{equation(1)}
	MinI=2427070.6150 + 3^{\textrm{d}}.7863886\times{E},
\end{eqnarray}
and the values of O-C for all eclipsing times are computed using equation (\ref{equation(1)}). As shown in Fig. \ref{fig:figure2}, the O-C values show a linear distribution, indicating that the orbital period needs to be revised.
\begin{table*}
	\begin{center}
		\footnotesize
		\caption{Eclipsing times of RX Dra.This table is available in full in machine-readable form. \label{tab:tab 1}}
		\begin{tabular}{lllllc}\hline
			Eclipsing times (P) &Errors&Eclipsing times (S) &Errors&Source\\
			(HJD)&&(HJD)\\\hline
			2458687.02119 &  0.00020 & 2458685.12819 &  0.00039& TESS\\
			2458690.80819 &  0.00022 & 2458688.91519 &  0.00042& TESS\\
			2458694.59419 &  0.00021 & 2458692.70119 &  0.00047& TESS\\
			2458698.38119 &  0.00020 & 2458700.27419 &  0.00044& TESS\\
			
			............... & ......... & ............... & .........&..\\ 
			2417502.400 & 0.003 & 2417689.859 & .........& O-C gateway\\
			2417551.631 & 0.002 & 2418132.829 & .........& O-C gateway\\
			............... & ......... & ............... & .........&..\\
			\hline
		\end{tabular}
	\end{center}
\end{table*}
After correcting for all eclipsing times using the linear fit method, a better linear ephemeris is
\begin{eqnarray}\label{equation(2)}
	MinI=2427070.6195 (\pm0.0011) + 3^{\textrm{d}}.7863954 (\pm 0.0000083)\times{E},
\end{eqnarray}
and the corresponding residuals are shown in the lower panel of Fig. \ref{fig:figure2} which are distributed close to the horizontal line (the red line) of zero with no visible trend of variation.

\section{Photometric model}
\label{W-D}
As shown in the lower panel of Fig. \ref{fig:figure1}, the two peaks of the TESS light curve are at the same level near TJD 2390. We therefore chose a sector (s40) of the light curve of about 30 days to analyze the basic mode solution without the stellar spot. In order to produce an average light curve to remove the effect of brightness pulsation as far as possible, the following two steps are taken: (i) TJD is converted to phase, and (ii) data points are reduced from approximately 19018 to 500 by averaging (i.e. points within 0.002 phase are averaged as one point). Finally, this light curve was analyzed using the Wilson-Devinney (W-D) program (\citealp{1971ApJ...166..605W,1979ApJ...234.1054W,1990ApJ...356..613W,2007ApJ...661.1129V,2012AJ....144...73W}).
\begin{table*}
	\begin{center}
		\footnotesize
		\caption{Photometric solutions for RX Dra of Model 2 using the W-D method.\label{tab:tab 2}}
		\begin{tabular}{llcc}\hline
			Parameters &&Model 2\\\hline
			$NF^a$&&90$\times$90\\
			$g_1$$^a$&&0.32\\
			$g_2$$^a$&&0.32\\
			$A_1$$^a$&&0.50\\
			$A_2$ &&0.204 (20)\\
			$T^a_1$(K) &&6987\\
			$i$ ($^o$) &&88.73 (12)\\
			$q$ ($M_2/M_1$)&&0.9026 (32)\\
			$T_2/T_1$&&0.9320(34)\\
			$L_1/(L_1+L_2)_{TESS}$&&0.7080(81)\\
			$L_2/(L_1+L_2)_{TESS}$&&0.2919(81)\\
			$R_2/R_1$&&0.715(74)\\
			$\Omega_1$&&8.7568(55)\\
			$\Omega_2$&&10.941(33)\\
			$r_1(pole)^b$&&0.1272(92)\\
			$r_2(pole)^b$&&0.0912(34)\\
			$r_1(point)^b$&&0.1278(95)\\
			$r_2(point)^b$&&0.0913(34)\\
			$r_1(side)^b$&&0.1274(93)\\
			$r_2(side)^b$&&0.0912(34)\\
			$r_1(back)^b$&&0.1277(94)\\
			$r_2(back)^b$&& 0.0913(34)\\
			mean residuals &&0.00037\\
			\hline
		\end{tabular}
	\end{center}
	\footnotesize {$^{a}$ These are obtained by the method described in this section}.\\
	\footnotesize {$^{b}$ In units of semi-major axis}.\\
	\footnotesize {The numbers in brackets are the errors in the last two bits of the data. The units of most of the parameters are dimensionless, except for those already mentioned}.\\
\end{table*}

Before we can model the light curve, we need to set up the following necessary parameters. The surface temperature of the primary star is 6987 K, which is obtained from Gaia's DR2 database \citep{2016A&A...595A...1G,2018A&A...616A...1G}. According to the spectral type given by Gaia, the primary star of RX Dra is an F-type star, and its bolometric albedo A$_{1}$ and gravity-darkening coefficient g$_{1}$ can therefore be assumed to be 0.50 \citep{2001MNRAS.327..989C} and 0.32 \citep{1967ZA.....65...89L}, respectively. Since the detailed spectral information of the secondary star is unknown, so we assumed that it is an G or K type star and set its bolometric albedo A$_{2}$ and gravity-darkening coefficient g$_{2}$ as the same of the primary. Linear limb darkening coefficients $x_{1}$ and $x_{2}$ for the primary and the secondary are from tables in \citealp{1993AJ....106.2096V}. The light curves in the Fig.\ref{fig:figure1} show that RX Dra is a binary system of the EA type, we choose to model its light curve using models 2 and 5 of the W-D code, where the grid fineness for the micro-integration on each surface element $NF$ is set by default to a more precise value of 90 $\times$ 90. Finally, after continuous iterations, we had a convergent solution in Model 2 by adjustment of the following variable parameters. These include the mass ratio $q$ $(M_2/M_1)$, the luminosity $L{_1}$ of star 1, the orbital inclination $i$, the dimensionless potential $\Omega_1$ for star 1 and $\Omega_2$ for star 2, the bolometric albedo A$_2$ of star 2, and the temperature $T_2$ of star 2.

\begin{figure}
	\centering 
	\includegraphics[width=0.85\textwidth]{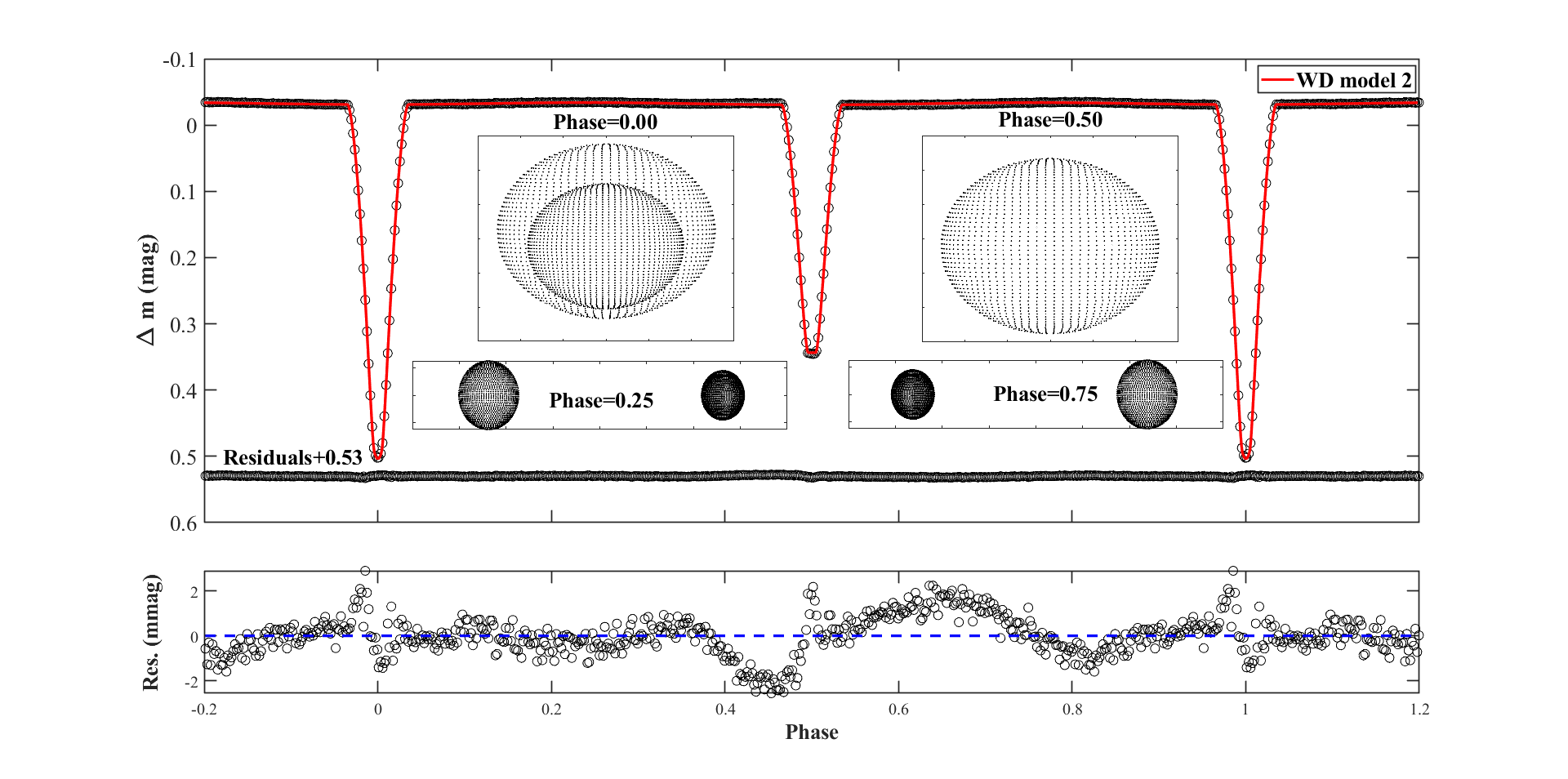}	
	\caption{Upper panel: the average light curve (the black circles) and the calculated fitting light curves (the red solid line), as well as the W-D theoretical geometric structures of RX Dra at phases 0.00, 0.25, 0.50 and 0.75. Lower panel: the enlarged view of residuals, which also shows some pulsations.} 
	\label{fig:figure3}%
\end{figure}

The parameters of the convergent solution of model 2 are listed in the Table \ref{tab:tab 2},  where the mass ratio $q$ is estimated to be 0.9026 ($\pm$0.0032), which means the masses of two components are close to each other. The primary star contributes about 71 percent of the luminosity of the entire system. 

The top panel of Fig. \ref{fig:figure3} shows the corresponding plots of the results based on model 2, which includes the mean curve (the black circle points), the residual curves (the black circle points), the fitted curves (the red solid lines) and the geometric structures of the binary system. The geometric structures are a visible indication that the primary star is larger than the secondary component and that there is a large separation between them. In addition, in combination with the relative theory of Roche's lobe that had been incorporated into the W-D code, we also knew that the volume fill factors ($V_{star}/V_L$) of the primary and secondary components were 3.55 ($\pm$0.94)$\%$ and 1.50 ($\pm$0.96)$\%$, respectively. So we proposed that the binary system, which could be made up of two main-sequence stars, is detached. 

As shown in Fig. \ref{fig:figure3}, the secondary eclipse is close to 0.5 phase, indicating that the two components are moving in a circular orbit around each other. One possible explanation for this is that the two components are in a tidally locked and synchronous rotational state. So we set the orbital eccentricity $e$ and the dimensionless rotation parameters $F1$ and $F2$ for star1 and star2 to 0 in the W-D code. The convergent results we obtained support our hypothesis.

\section{Stellar Model and Asteroseismic Analysis for RX Dra } \label{sec:stellar model}

\cite{2022MNRAS.515.2755S} had carried out an asteroseismic analysis for RX Dra. Their work is excellent, but they did not do any mode identification of the pulsation frequencies they have obtained. Table \ref{tab:2} shwows the 24 frequencies of them. In the section \ref{sec:photometric model}, the analysis of the photometric model shows that the primary star of RX Dra is more massive and contributes about 70 $\%$ luminosity to the system, so we assume that the primary component is a pulsating star and such frequencies from this star. We used the stellar model to perform further an asteroseismic analysis based on their work and to estimated the physical properties of the primay star.
\begin{table*}\scriptsize
	\begin{center}
		\caption{The $\gamma$ Dor pulsating frequencies of RX Dra, which had been obtained by \cite{2022MNRAS.515.2755S}.\label{tab:2}}
		\begin{tabular}{llccllccllcc}\hline
			ID&&Frequency &&Amplitude&&Phase&&Combination&&S/N\\
			&&(d$^{-1}$)&&(mmag) &&(rad)\\\hline
			F1&&0.40386 (2)&&0.407 (5) &&0.104 (2)&&-&&09.4\\
			F2&&0.48268 (1)&&0.64 (5) && -0.486 (1)&&-&& 15.2\\
			F3&&0.51792 (3)&&0.252 (5) &&-0.484 (3)&&F19-F11&& 06.0\\
			F4&&0.54059 (2)&&0.376 (5) &&0.296 (2)&&-&&09.1\\
			F5&&0.56659 (2)&&0.516 (5) &&0.492 (2)&&-&&12.5\\
			F6&&0.58234 (2)&&0.395 (5) &&0.371 (2)&&-&& 09.6\\
			F7&&0.63871 (2)&&0.41 (5) &&0.235 (2)&&&& 10.1\\
			F8&&0.65485 (2)&&0.373 (5) &&0.491 (2)&&F17-F2&& 09.2\\
			F9&&0.69392 (1)&&0.596 (5) &&0.204 (1)&&F18-F2&& 14.8\\
			F10&&0.72459 (4)&&0.224 (5) &&0.401 (4)&&F20-F6&& 05.6\\
			F11&&0.75959 (1)&&0.674 (5) && -0.414 (1)&&-&& 17.0\\
			F12&&0.77293 (2)&&0.397 (5) && -0.096 (2)&&-&& 10.0\\
			F13&&0.81678 (2)&&0.365 (5) &&0.391 (2)&&-&& 09.3\\
			F14&&0.91214 (2)&&0.526 (5) && -0.26 (2)&&-&& 13.8\\
			F15&&0.92597 (2)&&0.393 (5)&&0.115 (2)&&-&& 10.4\\
			F16&&1.10811 (5)&&0.179 (5)&& -0.442 (5)&&-&& 05.0\\
			F17&&1.13749 (2)&&0.466 (5)&&0.479 (2)&&-&& 13.2\\
			F18&&1.176625 (6)&&1.405 (5)&& 0.2303 (6)&&-&& 41.0\\
			F19&&1.27764 (3)&&0.276 (5)&&0.265 (3)&&-&& 08.5\\
			F20&&1.30693 (3)&&0.274 (5)&& -0.002 (3)&&-&& 08.6\\
			F21&&1.5985 (6)&&0.132 (5)&& -0.03 (6)&&-&& 05.1\\
			F22&&1.71749 (4)&&0.187 (5)&&0.324 (4)&&-&& 07.7\\
			F23&&1.74289 (3)&&0.289 (5)&& -0.44 (3)&&-&& 12.1\\
			F24&&2.8467 (1)&&0.054 (5)&&0.37 (1)&&-&& 04.3\\
			\hline
		\end{tabular}
	\end{center}
\end{table*}
\subsection{Input Physics}
The stellar evolution code Modules for Experiments in Stellar Astrophysics (MESA) is developed by \cite{2011ApJS..192....3P,2013ApJS..208....4P,2015ApJS..220...15P,2018ApJS..234...34P},  which is used to compute evolutionary and pulsational models. In particular, the sub-module ``pulse-adipls" in version r22.11.1 is used for the generation of stellar evolution models. In particular, to generate stellar evolution models and calculate the adiabatic frequencies of their non-radial g-modes \citep{2008Ap&SS.316..113C}, the submodule ``pulse-adipls" in version r12.11.1 is used. Our work uses the 2005 update of the OPAL equation of state tables \citep{2002ApJ...576.1064R}. The OPAL opacity tables from \cite{1996ApJ...464..943I} are used for the high temperature regions and the tables from \cite{2005ApJ...623..585F} are used for the low temperature regions. The assumption is that the initial component of metallicity is identical to that of the Sun \citep{2009ARA&A..47..481A}. The classical mixing length theory of \cite{1958ZA.....46..108B} with $\alpha$=1.90 \citep{2011ApJS..192....3P} is used in the convective zone. We adopt an exponentially decaying prescription and introduce an overshooting mixing diffusion coefficient \citep{1996A&A...313..497F, 2000A&A...360..952H} for the overshooting mixing of the convective core
\begin{eqnarray}\label{equation (1)}
	\mathit{D}_{\rm ov}=\mathit{D}_{\rm 0} \exp (\frac{-2z}{\mathit{f}_{\rm ov} \mathit{H}_{\rm p}}).
\end{eqnarray}
where $\mathit{D}_{\rm 0}$ indicates the diffusion mixing coefficient near the edge of the convective
core, $z$ presented the distance into the radiative zone away from the edge, $\mathit{f}_{\rm ov}$ are adjustable parameters describing the efficiency of the overshooting mixing, and $\mathit{H}_{\rm p}$ is the pressure scale height. The lower limit of the diffusion coefficient is set to $\mathit{D}^{\rm limit}_{\rm ov}$ = 1$\times$ 10$^{-2}$ cm$^2$ s$^{-1}$ \citep{2019ApJ...887..253C}, below which the overshoot is turned off in our work. In addition, elemental diffusion and magnetic fields on stellar structure and evolution are not included in this work.

\subsection{Grid of Stellar Models}
The evolutionary path and internal structure of a star depend on the initial stellar mass $M$, the initial chemical composition $(X,Y,Z)$ and the overshooting parameter $\mathit{f}_{\rm ov}$. Following the method of \cite{2018MNRAS.475..981L}, the initial helium abundance is set to be $Y=0.249+1.33Z$, so the stellar evolution and structure can be completely determined by the values of $M$,$Z$ and $\mathit{f}_{\rm ov}$. We considered the grid search of stellar masses $M$ between 1.50 M$_{\odot}$ and 1.80 M$_{\odot}$ with a step of 0.01 M$_{\odot}$, and the fraction of metallicity $Z$ is set to a range of 0.008 to 0.018 with steps of 0.001. The intermediate overshooting ($\mathit{f}_{\rm ov}$=0.02), reported by \cite{2019ApJ...887..253C}, is adopted to process overshooting mixing.  According to the effective temperature, the gravitational acceleration, and the luminosity values of Gaia DR2 and DR3 \citep{2016A&A...595A...1G,2018A&A...616A...1G}, and considering the spectral type and luminosity for the $\gamma$ Dor pulsating stars, the effective temperature is set to be 6500 K $< T_{eff} <$ 7500 K, the gravitational acceleration is set as 3.98 cms$^{-2}$$<$ log g $<$ 4.08 cms$^{-2}$, and the luminosity is set as 0.97 L$_{\odot} <log L <$ 1.07 L$_{\odot} $. Each star in the grid is calculated from the zero-age main-sequence to the post-main-sequence stage.

In addition, we also considered the rotation period $\mathit{P}_{\rm rot}$ between 3 days to 10 days with a step of 0.1 days, as the third adjustable parameter. Each pulsation mode will be split into 2 $l$+1 different frequencies for a given period $P_{rot}$ according to
\begin{eqnarray}\label{equation(4)}
	\mathit{\nu}_{l,n,m}=\mathit{\nu}_{l,n}+m \mathit{\delta\nu}_{l,n}=\mathit{\nu}_{l,n}+\mathit{\beta}_{l,n}\frac{m}{\mathit{P}_{rot}},
\end{eqnarray}
(Saio 1981; Dziembowski \& Goode 1992; Aerts et al. 2010), where $\mathit{\delta\nu}_{l,n}$ is splitting frequency, $ \mathit{\beta}_{l,n}$ is the rotational parameter which determines the size of rotational splitting. In the book of Aerts et al. 2010, the general expression of $ \mathit{\beta}_{l,n}$ of a uniformly rotating star is a simple equation
\begin{eqnarray}\label{equation(5)}
	\mathit{\beta}_{l,n}=\frac{\int_{0}^{R}(\mathit{\xi}^2_{r}+L^2\mathit{\xi}^2_{h}-2\mathit{\xi}_{r}\mathit{\xi}_{h}-\mathit{\xi}^2_{h}) r^2 \rho dr}{\int_{0}^{R}(\mathit{\xi}^2_{r}+L^2\mathit{\xi}^2_{h}) r^2 \rho dr},
\end{eqnarray}
where $\mathit{\xi}_{r}$ and $\mathit{\xi}_{h}$ describe the displacement in radial and horizontal directions, $\rho$ is local density, and $L^2=l(l+1)$.

\subsection{Fitting Results of RX Dra}
We compare model frequencies with observed frequencies to find the optimal model for the observations in Table \ref{tab:tab 3}, except for the combination frequencies $F3$, $F8$, $F9$ and $F10$
\begin{eqnarray}\label{equation(6)}
	S^2=\frac{1}{k} \Sigma(|f_{model, i}-f_{obs, i}|^2),
\end{eqnarray}
where $f_{obs, i}$ indicates the observed frequency, $f_{model, i}$ represents its corresponding model frequency, and $k$ is the number of observed frequencies. The independent frequencies in Table \ref{tab:tab 3} are not identified in advanced, so the theoretical frequencies closest to them are taken as their possible model counterparts.
\begin{figure}
	\centering 
	\includegraphics[width=0.8\textwidth]{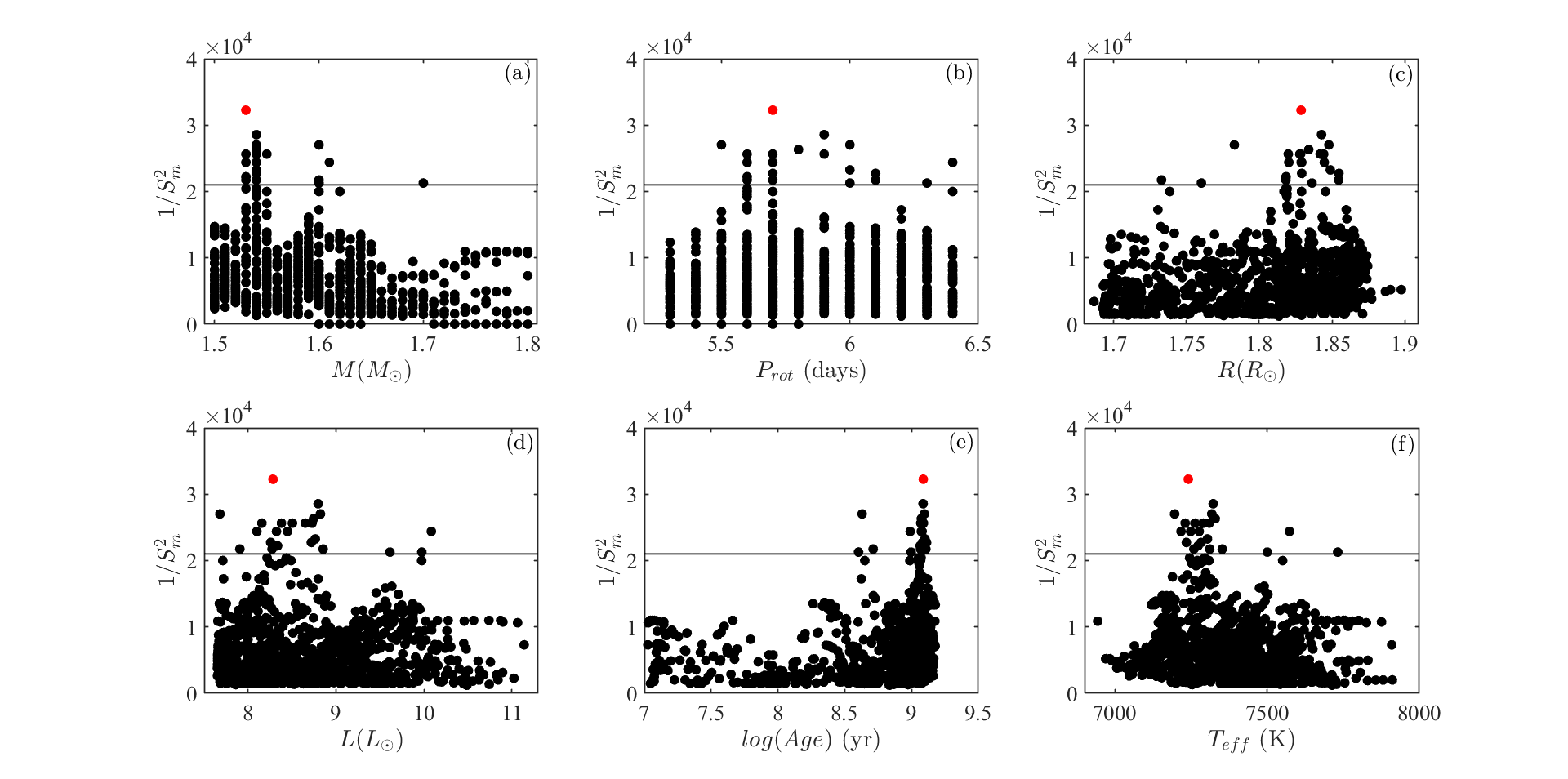}	
	\caption{Visualization of fitting results $S^{2}_m$ and physical parameters: the stellar mass $M$, the rotation period $\mathit{P}_{rot}$, the radius $R$, the luminosity $L$, the age $\log (Age)$, and the effective temperature $\mathit{T}_{\rm eff}$, respectively. The horizontal line in black marks the position of $S^{2}_m$ = 0.0000476. The black dots indicate the minimum chi2 $S^{2}_m$ of each model and the red filled circles indicate the minimum chi2 $S^{2}_m$ of the best-fitting model.} 
	\label{fig:figure4}%
\end{figure}
\begin{figure}
	\centering 
	\includegraphics[width=0.8\textwidth]{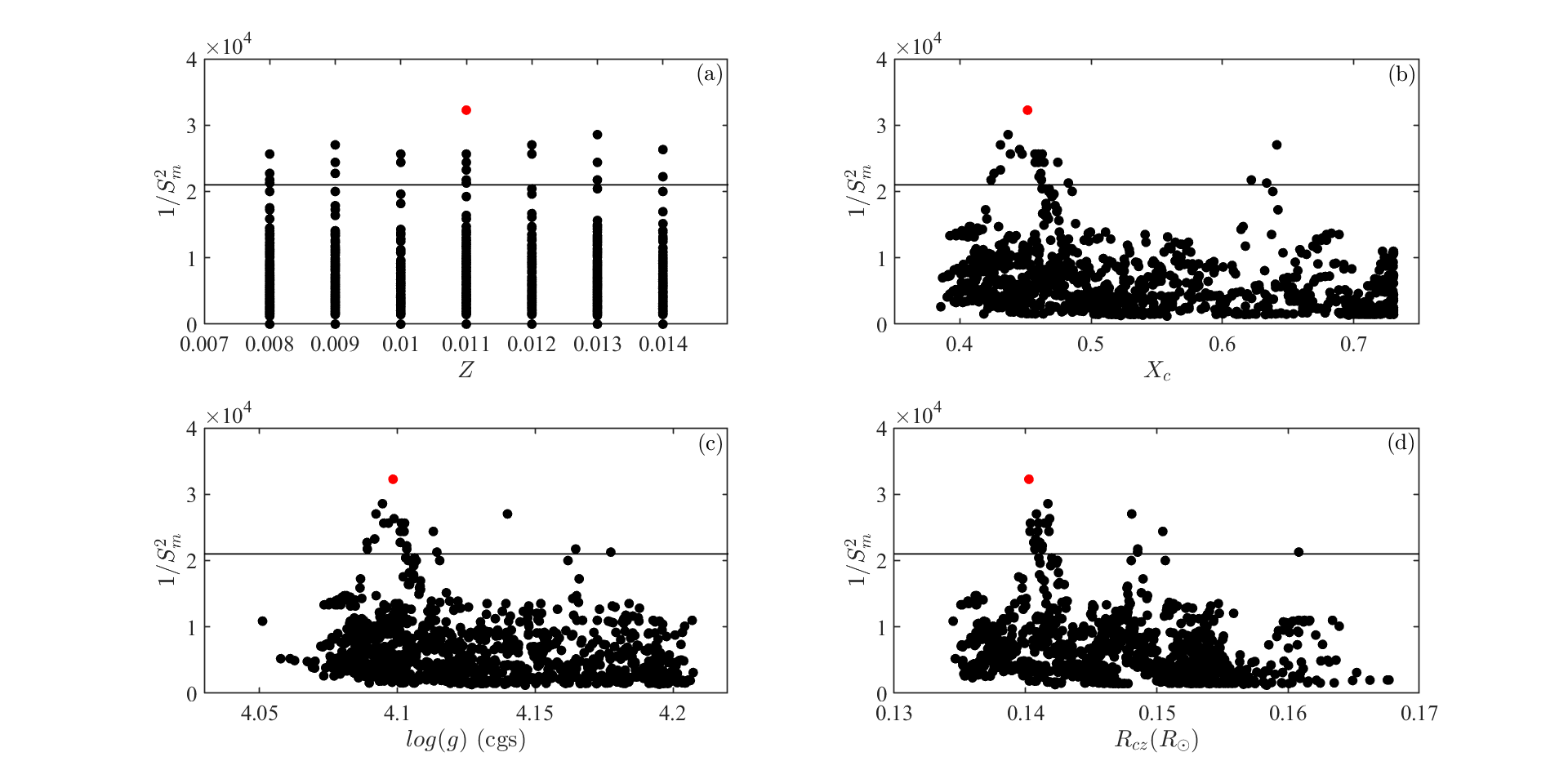}	
	\caption{Visualization of fitting results $S^{2}_m$ and physical parameters: the fraction of metallicity $Z$, the fraction of center hydrogen $\mathit{X}_{\rm c}$, the gravitational acceleration $\log (g)$, and the radius of convective core $\mathit{R}_{\rm cz}$, respectively. The horizontal line in black marks the position of $S^{2}_m$ = 0.0000476. The black dots indicate the minimum chi2 $S^{2}_m$ of each model and the red filled circles indicate the minimum chi2  $S^{2}_m$ of the best-fitting model.} 
	\label{fig:figure5}%
\end{figure}

The relation of resulting $S^{2}_m$ versus various physical parameters are shown in Fig. \ref{fig:figure4} and Fig. \ref{fig:figure5}. The horizontal line in black marks the position of $S^{2}_m$ = 0.0000476. The black dots above the horizontal line show the candidate models and they are listed in Table \ref{tab:tab 4} and the red filled circles indicate the minimum chi2  $S^{2}_m$ of the best-fitting model (model 10 in Table \ref{tab:tab 4}).
\begin{table*}
	\begin{center}
		\caption{Candidates models with $S^{2}_m <$ 0.0000476.\label{tab:tab 4}}
		\begin{tabular}{llccllccllccc}\hline
			Mode ID&$X_{c}$& $M$&$\mathit{T}_{\rm eff}$&$\log(g)$&$R$&$L$&$\log(age)$&$P_{rot}$&$Z$&$\mathit{R}_{\rm cz}$&$S^{2}_m$\\ &&(M$_{\odot}$)&(K)&(cgs)&(R$_{\odot}$)&(L$_{\odot}$)&($\times$10$^{9}$yr)&(days)\\ \hline
			1	&0.4473	&1.55	&7289	&4.0967	&1.8442	&8.6502	&9.088	&5.9	&0.008	&0.1414	&0.000039\\
			2	&0.4615	&1.54	&7234	&4.1010	&1.8292	&8.2574	&9.0815	&5.7	&0.008	&0.1410	&0.000044\\
			3	&0.6220	&1.60	&7352	&4.1646	&1.7329	&7.906	&8.7129	&5.7	&0.008	&0.1485	&0.000046\\
			4	&0.6336	&1.70   &7732	&4.1773	&1.7602	&9.9759	&8.6032	&6.0	&0.008	&0.1608	&0.000047\\
			5	&0.6414	&1.60	&7196	&4.1399	&1.7829	&7.6794	&8.6302	&5.5	&0.009	&0.1480	&0.000037\\
			6	&0.4597	&1.54	&7263	&4.1014	&1.8283	&8.3790	&9.0751	&5.7	&0.010	&0.1409	&0.000039\\
			7	&0.4598	&1.54	&7249	&4.1010	&1.8292	&8.3225	&9.0791	&5.7	&0.009	&0.1409	&0.000041\\
			8	&0.4639	&1.53	&7217	&4.1024	&1.8203	&8.0992	&9.0782	&5.6	&0.010	&0.1403	&0.000041\\
			9	&0.4260	&1.54	&7283	&4.0890	&1.8547	&8.7186	&9.1122	&6.1	&0.009	&0.1406	&0.000044\\
			10	&0.4515	&1.53	&7241	&4.0984	&1.8288	&8.2830	&9.0875	&5.7	&0.011	&0.1403	&0.000031\\
			11	&0.4309	&1.54	&7318	&4.0922	&1.8478	&8.8229	&9.0959	&6.0	&0.012	&0.1408	&0.000037\\
			12	&0.4626	&1.53	&7231	&4.1026	&1.8199	&8.1566	&9.0748	&5.6	&0.011	&0.1403	&0.000039\\
			13	&0.4571	&1.54	&7291	&4.1016	&1.8278	&8.5043	&9.0691	&5.7	&0.012	&0.1414	&0.000039\\
			14	&0.4384	&1.54	&7310	&4.0950	&1.8418	&8.7273	&9.0886	&5.9	&0.012	&0.1416	&0.000039\\
			15	&0.4573	&1.54	&7277	&4.1012	&1.8288	&8.4482	&9.0732	&5.7	&0.011	&0.1417	&0.000041\\
			16	&0.4309	&1.54	&7304	&4.0918	&1.8488	&8.7635	&9.0999	&6.0	&0.011	&0.1407	&0.000043\\
			17	&0.4237	&1.54	&7311	&4.0891	&1.8545	&8.8535	&9.1065	&6.1	&0.011	&0.1407	&0.000046\\
			18	&0.4826	&1.60	&7500	&4.1143	&1.8362	&9.6133	&8.9966	&6.3	&0.011	&0.1485	&0.000047\\
			19	&0.4367	&1.54	&7323	&4.0946	&1.8428	&8.7971	&9.0863	&5.9	&0.013	&0.1417	&0.000035\\
			20	&0.4454	&1.54	&7329	&4.0988	&1.8339	&8.7451	&9.0729	&5.8	&0.014	&0.1418	&0.000038\\
			21	&0.4745	&1.61	&7573	&4.1130	&1.8446	&10.084&8.9898	&6.4	&0.013	&0.1504	&0.000041\\
			22	&0.4603	&1.53	&7273	&4.1033	&1.8184	&8.3352	&9.0648	&5.6	&0.014	&0.1412	&0.000045\\
			23	&0.4621	&1.53	&7259	&4.1034	&1.8182	&8.2722	&9.0675	&5.6	&0.013	&0.1412	&0.000046\\
			\hline
		\end{tabular}
	\end{center}
\end{table*}
The panels (a)-(c) of Fig. \ref{fig:figure4} present the changes of $S^{2}_m$ as a function of the stellar mass $M$, the rotation period $\mathit{P}_{\rm rot}$, and the stellar radius $R$. The mass $M$ converge to 1.53-1.70 M$_{\odot}$, the period $\mathit{P}_{\rm rot}$ distribute in a range of 5.5-6.4 days, and the radius converge to 1.7329-1.8547 R$_{\odot}$, respectively. And the panels (d)-(f) of Fig. \ref{fig:figure4} present the changes of $S^{2}_m$ as a function of the stellar luminosity $L$, the age $\log (Age)$ and the effective temperature $\mathit{T}_{\rm eff}$. The stellar luminosity shows a wide distribution between 7.6794-10.0845 L$_{\odot}$, while the age and the effective temperature display a narrow distribution between 8.6033-9.1122 years and 7196-7731 K.

The panels (a)-(b) of Fig. \ref{fig:figure5} present the changes of $S^{2}_m$ as a function of the fraction of metallicity $Z$, the fraction of center hydrogen $\mathit{X}_{\rm c}$. The fraction of metallicity have a good convergence between 0.008-0.014, and the center hydrogen distribute between 0.4237-0.6414. The panels (c)-(d) of Fig. \ref{fig:figure5} present the changes of $S^{2}_m$ as a function of the gravitational acceleration $\log(g)$ and the radius of convective core $\mathit{R}_{cz}$. They distribute between 4.0890-4.1774 cms$^{-2}$ and 0.1403-0.1608 R$_{\odot}$.

The fundamental parameters of the primary star derived from the asteroseismic models are listed in Table \ref{tab:tab 5} based on the above calculations.
\begin{table*}
	\begin{center}
		\setlength{\tabcolsep}{4pt}
		\renewcommand{\arraystretch}{1.2}
		\caption{Fundamental parameters of the primary star of RX Dra. The data in brackets are the values of the best fitting model, the upper and lower uncertainties are calculated using these values minus the maximum and minimum values of the candidate models.}\label{tab:tab 5}
		\begin{tabular}{llcc}\hline
			Parameters &&values\\\hline
			$M$ (M$_{\odot}$)&&1.53-1.70 (1.53$^{+0.00}_{-0.17}$)\\
			$\mathit{P}_{\rm rot}$ (days)&&5.5-6.4 (5.70$^{+0.7}_{-0.2}$)\\
			$\mathit{T}_{\rm eff}$ (K)&&7196-7731 (7240$^{+490}_{-44}$)\\
			$log g$ (cgs)&&4.0890-4.1774 (4.0984$^{+0.0784}_{-0.0094}$)\\
			$R$ (R$_{\odot}$)&&1.7329-1.8547 (1.8288$^{+0.0260}_{-0.0959}$)\\
			$L$ (L$_{\odot}$)&&7.6794-10.0845 (8.2830$^{+1.8015}_{-0.6036}$)\\
			$\mathit{X}_{\rm c}$&&0.4237-0.6414 (0.4515$^{+0.1899}_{-0.0278}$)\\
			$log (Age)$ (yr)&&8.6033-9.1122 (9.0876$^{+0.0246}_{-0.4843}$)\\
			$Z$ &&0.008-0.014 (0.011$^{+0.003}_{-0.003}$)\\
			$\mathit{R}_{\rm cz}$ &&0.1403-0.1608 (0.1403$^{+0.0206}_{-0.0000}$)\\
			\hline
		\end{tabular}
	\end{center}
\end{table*}
The parameters obtained from the asteroseismic models always agree with the parameters derived from the eclipsing light curve fit, such as CoRoT 100866999 \citep{2019ApJ...887..253C} and KIC 10736223 \citep{,2020ApJ...895..136C}.

\begin{table*}
	\begin{center}
		\caption{Comparison between model frequencies of the best-fitting model and the observed $\gamma$ Dor frequencies.\label{tab:tab 6}}
		\begin{tabular}{llccllccll}\hline
			ID&&$\mathit{F}_{\rm ob}$ ($\mu$Hz) &&$\mathit{F}_{\rm mod}$ ($\mu$Hz) &&$(l,n,m)$&$\mathit{\beta}_{\rm l,n}$&$|$$\mathit{F}_{\rm ob}$ - $\mathit{F}_{\rm mod}$$|$\\\hline
			F1&&4.67430&&4.67350&&(1,-72,0)&0.50&0.00701\\
			F2&&5.58657&&5.56920&&(1,-59,0)&0.50&0.01740\\
			F4&&6.25682&&6.28541&&(1,-53,0)&0.50&0.02861\\
			F5&&6.55775&&6.50248&&(1,-50,0)&0.50&0.05500\\
			F6&&6.74004&&6.76157&&(1,-49,0)&0.50&0.02151\\
			F7&&7.39247&&7.36894&&(1,-45,0)&0.50&0.02353\\
			F11&&8.79155&&8.85296&&(1,-37,0)&0.50&0.06140\\
			F12&&8.94594&&8.10105&&(1,-36,1)&0.50&0.84481\\
			F13&&9.45347&&9.36258&&(1,-34,-1)&0.50&0.09080\\
			F14&&10.55717&&10.45366&&(1,-31,-1)&0.50&0.10351\\
			F15&&10.71724&&10.72799&&(1,-30,0)&0.50&0.01075\\
			F16&&12.82534&&12.77079&&(1,-25,0)&0.50&0.05455\\
			F17&&13.16539&&13.25690&&(1,-24,1)&0.50&0.09151\\
			F18&&13.61834&&13.68172&&(1,-23,0)&0.50&0.06338\\
			F19&&14.78750&&14.66792&&(1,-21,-1)&0.50&0.11958\\
			F20&&15.12650&&15.34657&&(1,-20,1)&0.50&0.22007\\
			F21&&19.87835&&19.76264&&(2,-28,-1)&0.83&0.11571\\
			F22&&20.17233&&20.47411&&(2,-27,2)&0.83&0.30178\\
			F23&&18.50115&&18.40773&&(3,-43,-1)&0.91&0.09342\\
			F24&&32.94791&&32.00233&&(3,-23,0)&0.91&0.94558\\
			\hline
		\end{tabular}
	\end{center}
\end{table*}
The theoretical model frequencies of the best-fitting model are listed in Table \ref{tab:tab 6}. This table also shows comparisons between model frequencies of the best-fitting model and the observed $\gamma$ Dor frequencies. There are 16 frequencies (F1-F7, F11-F20) identified as dipole modes, two frequencies (F21, F22) identified as quadrupole modes, and another two frequencies (F23, F24) identified as sextupole modes, as shown in this table.

\section{Discussions and Conclusions}
In section \ref{W-D}, we have investigated the photometric model using the TESS light curve of RX Dra, and derived the orbital parameters of this binary system, which are listed in Table \ref{tab:tab 2}. In section \ref{MESA}, we presented our fitting results of asteroseismic model for the primary star of RX Dra. Some physical parameters of the theoretical models show a certain dispersion, while another shows a good convergence. On the basis of these parameters, we discussed the physical properties of this binary system.
\begin{figure}
	\centering 
	\includegraphics[width=0.6\textwidth]{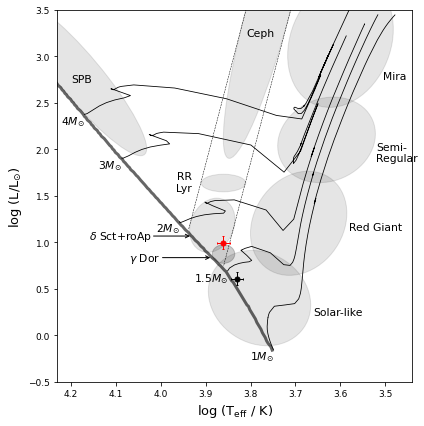}	
	\caption{The positions of the primary star (red dot) and the secondary component (black dot) in the H-R diagram.} 
	\label{fig:figure 6}%
\end{figure}

The following parameters can be calculated by entering the mass of the primary component into the W-D code; the absolute bolometric magnitude $\mathit{M}_{\rm bol1}$ = 2.56 ($\pm$0.07) mag and $\mathit{M}_{\rm bol2}$ = 3.59 ($\pm$0.07) mag. So the visual magnitude in V-band m$_V$ is calculated by (Chen et al. 2018)
\begin{eqnarray}\label{equation(7)}
	\log(L/L_{\odot})=0.4\times(4.74-\mathit{M}_{\rm bol}),
\end{eqnarray}
\begin{eqnarray}\label{equation(8)}
	\mathit{M}_{\rm V}=\mathit{M}_{\rm bol}-BC,
\end{eqnarray}
and
\begin{eqnarray}\label{equation(9)}
	\mathit{m}_{\rm V}=\mathit{M}_{\rm V}+5\times \log(1000/{\pi})-5+\mathit{A}_{\rm V},
\end{eqnarray}
where the parallax $\pi$ = 2.2185 ($\pm$0.0230) (Prusti et al. 2016; Brown et al. 2018), the interstellar extinction $\mathit{A}_{\rm V}$$\approx$ 0 and the bolometric correction $BC$ = -0.01 (Bessell et al. 1998). According to the equation (\ref{equation(7)}), the luminosity of the primary star is calculated as $L$= 7.40 ($\pm$ 0.48) L$_{\odot}$, which is close to the value of 8.2830 L$_{\odot}$ obtained by the MESA model. The maximum magnitude in V-band for HZ Dra is about 10.49 mag (Herwig 2000). Using the equations (\ref{equation(8)})-(\ref{equation(9)}), the calculation of the visual magnitude is $\mathit{m}_{\rm V}$ = 10.52 ($\pm$0.05) mag, which is agreement with the observational value 10.49. This  means that the photometric solutions and the MESA model are reliable and reasonable.

According to the results of our photomeric model and MESA model, the basic parameters of the secondary star can be calculated to be $M_2$= 1.38$^{+0.18}_{-0.00}$ M$_{\odot}$, $R_2$= 1.3075$^{+0.0450}_{-0.2543}$ R$_{\odot}$, $L_2$= 3.4145$^{+0.1320}_{-0.1843}$ L$_{\odot}$, 
$T_2$=6747$^{+201}_{-221}$ K. The distance between the two components thus is thus estimated by Kepler's third law to be $a$=14.60 $^{+0.47}_{-0.47}$ R$_{\odot}$. So we obtained the positions of the two components of RX Dra in the H-R diagram (Fig. \ref{fig:figure 6}) based on the above calculations, which shows that the primary star is a $\gamma$ Dor pulsator (the red dot), and the secondary star (the black dot) is in the position of a solar-like pulsator, indicating that this star could be a possible solar-like pulsating star. In other words, the binary system RX Dra should consist of two main-sequence stars, both pulsating.

\begin{figure}
	\centering 
	\includegraphics[width=1\textwidth, angle=0]{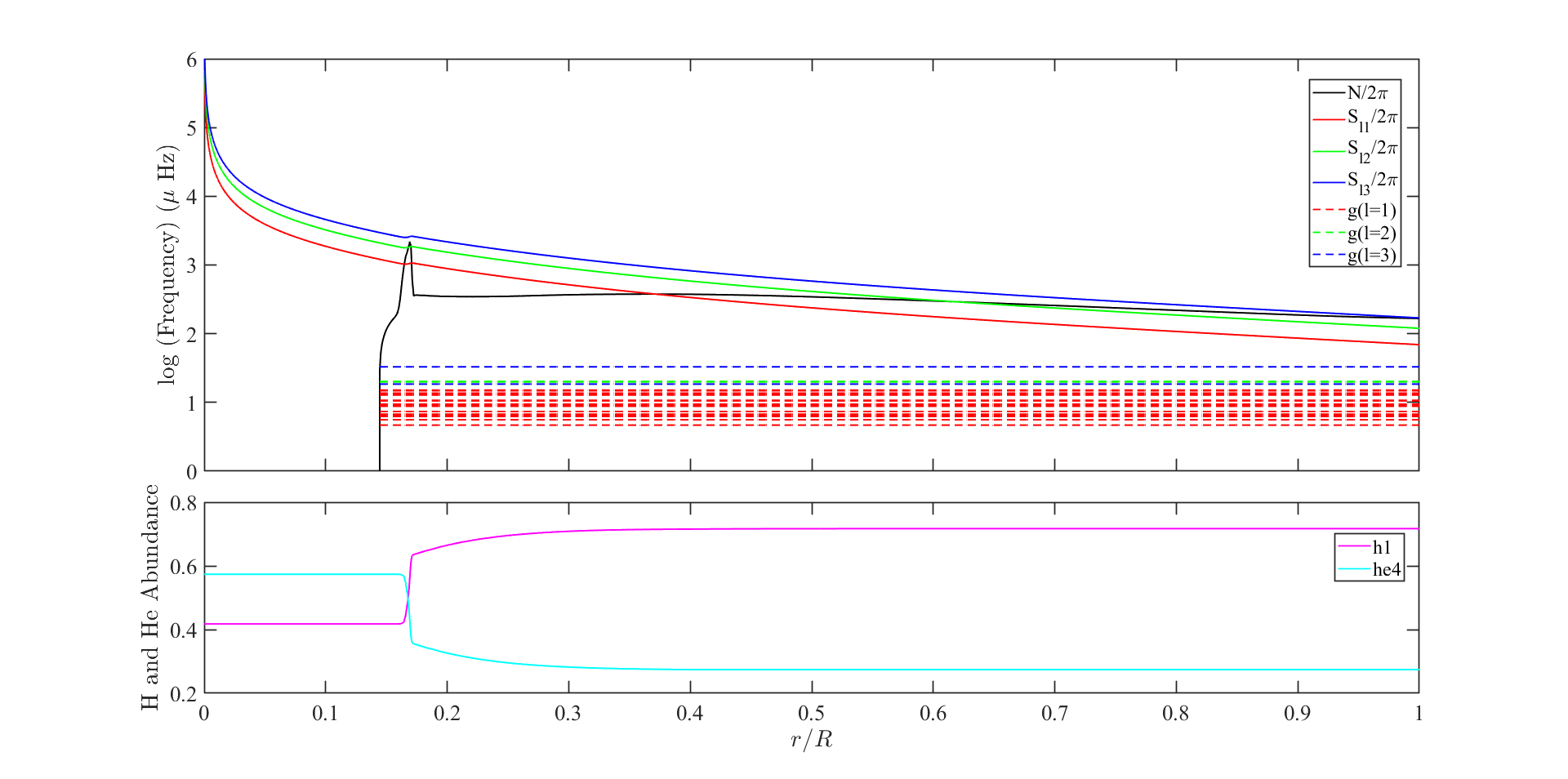}	
	\caption{Upper panel: the profile of the Brunt-V{\"a}is{\"a}l{\"a} frequency (the black line) and the characteristic acoustic frequencies $S_l$ ($l$=1,2,3) (the red, green, and blue lines). The propagation zone of the wave of the observed g-mode frequencies inside the primary star is marked by the red, green and blue dashed lines. Lower panel: the hydrogen (the pink line) and helium (the cyan line) abundances of the best-fitting model.} 
	\label{fig:figure 7}%
\end{figure}

The profiles of the Brunt-V$\rm \ddot{a}$is$\rm \ddot{a}$l$ \rm \ddot{a}$ frequency $N$, the characteristic acoustic frequencies $S_l$ ($l$=1,2,3) and the hydrogen (h1) and helium (he4) abundances within the best-fitting model for the primary star are shown in Fig. \ref{fig:figure 7}. This picture also shows that the Brunt-V$\rm \ddot{a}$is$\rm \ddot{a}$l$ \rm \ddot{a}$ frequency $N$ has a peak, while the hydrogen abundance has a maximum and the helium abundance a minimum at the same location, the corresponding wave of the observed g-mode can propagate in the regions satisfying $\omega < N$ and $\omega < S_l$ ($\omega$ is the circle frequency). The hydrogen of convective core will gradually be converted to helium as the star evolves over time. Meanwhile, the central convection will shrink into the interior of the core, causing a changing hydrogen abundance zone in the external edge of the convective core. Due to the elemental abundance gradient, the Brunt-V$\rm \ddot{a}$is$\rm \ddot{a}$l$ \rm \ddot{a}$ frequency $N$ has a maximum in this zone. We therefore obtain that the radius from the centre to the top of the convective core of the primary star, where the equation ($\nabla_r=\mathit{\nabla}_{\rm ad})$ is satisfied, is of the order of 0.1403$^{+0.0206}_{-0.0000}$ R $_{\odot}$.

Comparisons between model frequencies and observations suggest that the observed frequencies F1-F7 and F11-F20 are 16 high-order dipole g-modes, two frequencies (F21, F22) are high-order quadrupole g-modes, and frequencies (F23, F24) are two high-order sextupole g-modes. As can be seen in Table \ref{tab:tab 6}, six observed frequencies (F12, F14, F19, F20, F22 and F24) are not very close to the corresponding theoretical frequencies when compared with other frequencies. We think that this phenomenon may be caused by the pulsation of the secondary star, since it is a possible solar-like pulsator, but further studies are needed.

The rotation period of the primary star is determined to be $\mathit{P}_{\rm rot}$=5.7$^{+0.7}_{-0.2}$ days, which is slower than the orbital period $\mathit{P}_{\rm rot}$=3.78 days. The first order effect of the rotation on the pulsation is proportional to 1/$\mathit{P}_{\rm rot}$, and the second order effect is proportional to 1/($\mathit{P}^2_{\rm rot}$ $\nu_{l,n}$) (Saio 1981; Dziembowski et al. 1992, Aerts et al. 2010 ). Then the ratio of them can be estimated to be on the order 1/($\mathit{P}_{\rm rot}$ $\nu_{l,n}$), where the 1/$\mathit{P}_{\rm rot}$ is estimated to be 2.03 $\mu$Hz and $\nu_{l,n}$ ranges from 4.64519 to 32.93316 $\mu$Hz. The second effect of rotation on pulsation is much smaller than that of the first order. Therefore, it is not considered in our work.

\section{Acknowledgments}
This work is supported by the International Cooperation Projects of  (No. 2022YFE0127300) the National Key R\&D Program , 
the National Natural Science Foundation of China (No. 11933008), the Young Talent Project of  ``Yunnan Revitalization Talent Support Program" in Yunnan Province, the basic research project of Yunnan Province (Grant No. 202201AT070092), CAS ``Light of West China" Program.
This work has made use of data from the European Space Agency (ESA) mission Gaia. Processed by the Gaia Data Processing and Analysis Consortium.Funding for the DPAC has been provided by national institutions, in particular the institutions participating in the $Gaia$ Multilateral Agreement. The TESS data presented in this paper were obtained from the Mikulski Archive for Space Telescopes (MAST) at the Space Telescope Science Institute (STScI). STScI is operated by the Association of Universities for Research in Astronomy, Inc. Support to MAST for these data is provided by the NASA Office of Space Science. Funding for the TESS mission is provided by the NASA Explorer Program.


\bibliography{sample631}{}
\bibliographystyle{aasjournal}



\end{document}